\documentclass[aps,prl,groupedaddress,showpacs,twocolumn]{revtex4-1}
\usepackage[utf8]{inputenc}
\usepackage{hyperref}
\usepackage{graphicx}  
\usepackage{dcolumn}   
\usepackage{bm}        
\usepackage{amssymb,amsmath}   
\usepackage{uri}

\usepackage[x11names, rgb, html]{xcolor}
\definecolor{sbase03}{HTML}{002B36}
\definecolor{sbase02}{HTML}{073642}
\definecolor{sbase01}{HTML}{586E75}
\definecolor{sbase00}{HTML}{657B83}
\definecolor{sbase0}{HTML}{839496}
\definecolor{sbase1}{HTML}{93A1A1}
\definecolor{sbase2}{HTML}{EEE8D5}
\definecolor{sbase3}{HTML}{FDF6E3}
\definecolor{syellow}{HTML}{B58900}
\definecolor{sorange}{HTML}{CB4B16}
\definecolor{sred}{HTML}{DC322F}
\definecolor{smagenta}{HTML}{D33682}
\definecolor{sviolet}{HTML}{6C71C4}
\definecolor{sblue}{HTML}{268BD2}
\definecolor{scyan}{HTML}{2AA198}
\definecolor{sgreen}{HTML}{859900}
\hypersetup{
  colorlinks = true,
  allcolors = {blue}
}

\usepackage{color}

\begin{document}
\title{Arcsine Laws in Stochastic Thermodynamics}

\author{Andre C. Barato$^{1}$}\email{barato@pks.mpg.de}
\author{\'Edgar Rold\'an$^{1,2}$}\email{edgar@pks.mpg.de}
\author{Ignacio A. Mart\'inez$^{3}$}
\author{Simone Pigolotti$^4$}
\affiliation{\vspace{0.1cm}$^1$ Max Planck Institute for the Physics of Complex Systems, N\"othnizer Strasse 38, 01187 Dresden,Germany\vspace{0.01cm}\\
$^2$ The Abdus Salam International Centre for Theoretical Physics, Strada Costiera 11, 34151, Trieste, Italy\vspace{0.01cm}\\
$^3$ Departamento de Estructura de la Materia, F\'isica Termica y Electronica and GISC, Universidad Complutense de Madrid 28040 Madrid, Spain\vspace{0.01cm}\\
$^4$ Okinawa Institute of Science and Technology Graduate University, Onna, Okinawa 904-0495, Japan}

\begin{abstract}
  We show that the fraction of time a
  thermodynamic current spends above its average value follows the arcsine
  law, a prominent result obtained by L\'evy for Brownian motion. Stochastic currents with long streaks above or below their
  average are much more likely than those that spend similar fractions of
  time above and below their average. Our result is  confirmed with experimental data from a 
  Brownian Carnot engine. We also conjecture that two other random times associated with currents obey the arcsine law: the time a current reaches its maximum value and the last time a current crosses its average value.
  These results apply to, {\sl inter alia}, molecular motors, quantum dots and
colloidal systems.
\end{abstract}

\maketitle

In 1939,  Paul L\'{e}vy calculated the
distribution of the fraction of time $\mathcal{T}$ that a
trajectory of Brownian motion stays above zero~\cite{levy39}. L\'evy proved that this fraction of time  is distributed
according to
\begin{equation}
P(\mathcal{T})=
\frac{1}{\pi}\frac{1}{\sqrt{\mathcal{T}(1-\mathcal{T})}}\; .
\label{eq:Levy}
\end{equation}
This result and related extensions are often
referred to as the ``arcsine law'' \cite{erdo47,spit56,feller,maju05}.
The name stems from the fact that the cumulative distribution of
$\mathcal{T}$ reads
$F(\mathcal{T})=\int_0^\mathcal{T} P(\mathcal{T}')\text{d}\mathcal{T}'=
\frac{2}{\pi}\arcsin(\sqrt{\mathcal{T}})$.  A counterintuitive aspect
of the U-shaped distribution~\eqref{eq:Levy} is that its average value
$\langle\mathcal{T}\rangle=1/2$ corresponds to the minimum of the distribution,
i.e., the less probable outcome, whereas values close to the extrema
$\mathcal{T}=0$ and $\mathcal{T}=1$ are much more likely. 
Brownian trajectories with a long "winning" (positive) or "losing" (negative) streak are quite likely.

Several phenomena in physics and biology have been shown to be described by the arcsine law and related distributions. 
 Examples include conductance in disordered materials~\cite{nazarov1994limits,beenakker1997random}, chaotic dynamical systems~\cite{akimoto2008generalized}, partial melting of polymers~\cite{oshanin2009helix}, quantum chaotic scattering~\cite{mejia2011symmetry} and generalized fractional Brownian processes~\cite{sadhu2018generalized}. Notably, the arcsine law~\eqref{eq:Levy} has also been explored in finance
\cite{shiryaev2002quickest}, where investment strategies can lead to a
much smaller alternance of periods of gain and loss than one would
expect based on naive arguments.


Recent theory and experiments extended thermodynamics to mesoscopic
systems that are driven away from
equilibrium~\cite{bust05,seif12,parr15,peko15,proe16,mart17,cili17}. Mesoscopic systems
operate at energies comparable with the thermal energy $k_BT$, where
$k_B$ is the Boltzmann constant and $T$ is the temperature. At these
energy scales, observables such as work, heat, entropy production, and
other thermodynamic currents are not deterministic
as in macroscopic thermodynamics, but rather stochastic
quantities~\cite{seki98}.

\begin{figure}
\includegraphics[width=75mm]{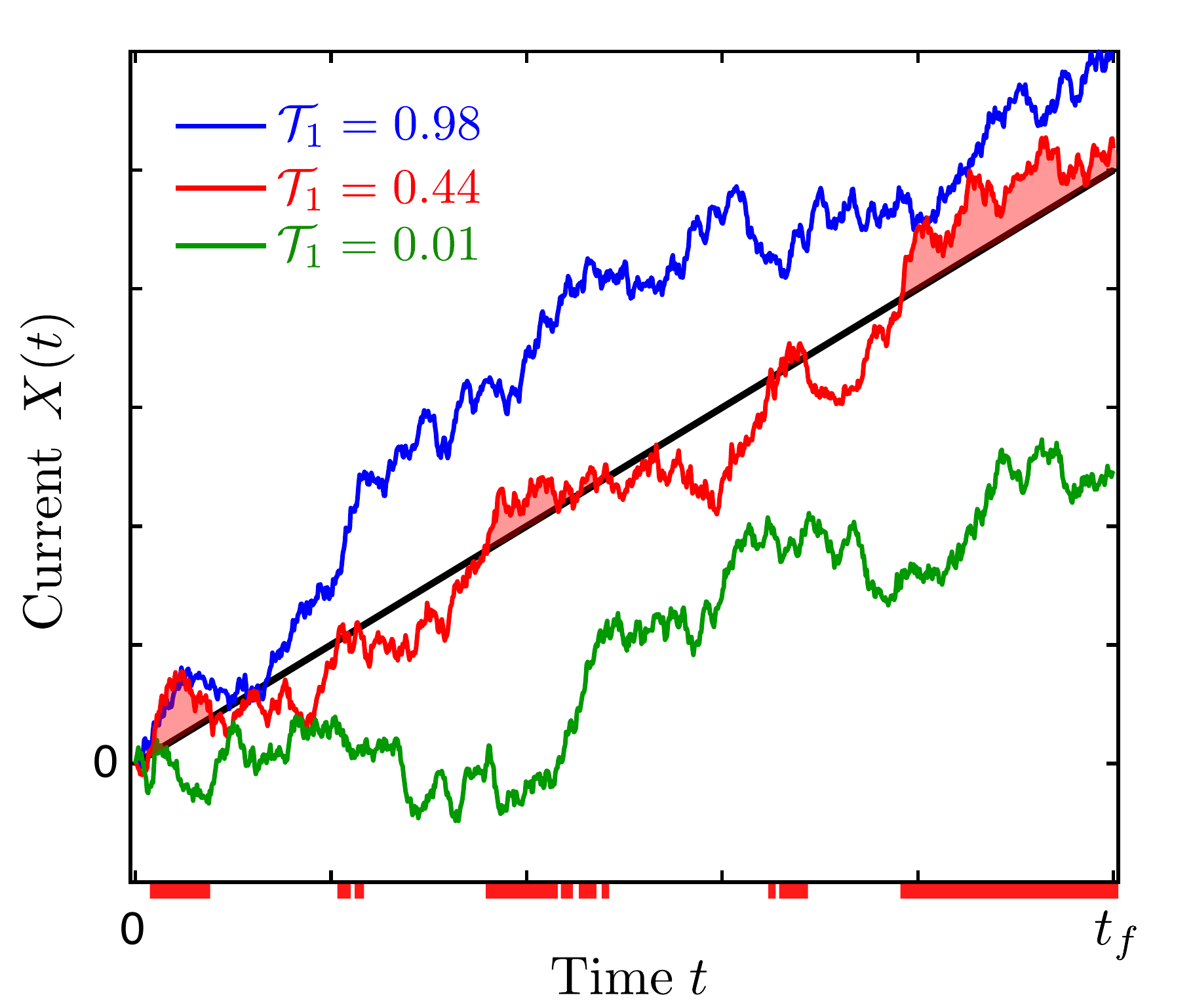}
\vspace{-2mm}
\caption{{\bf Illustration of the fraction of time $\mathcal{T}_1$ elapsed by a thermodynamic current above its average}.   The figure shows
  three different realizations of a stochastic current (colored lines)
  and its average linear growth (black line).
  For the example of the red trajectory, we mark in the x-axis the time intervals for
  which the trajectory stays above the average (red shaded areas).
  The values of $\mathcal{T}_1$  for
  the three trajectories are shown in the legend.}
\label{fig:1} 
\end{figure}

While the concept of a fluctuating entropy was already suggested by
the forefathers of thermodynamics and statistical
physics~\cite{maxw1878}, the universal statistical properties of
thermodynamic currents discovered in the last two decades 
have  extended thermodynamics, providing novel insights that  also apply to the nanoscale.
Prominent examples are fluctuation relations
\cite{boch79,gall95,jarz97,kurc98,lebo99,croo99,seif05a}, which generalize the second law of thermodynamics.  More
recently, several other universal results have been obtained. They include a
relation between precision and dissipation known as thermodynamic
uncertainty relation~\cite{bara15a,piet16,ging16}, stopping-time and
extreme-value distributions of entropy production (and related
observables)~\cite{sait16,neri17,pigo17,garr17}, and efficiency
statistics for mesoscopic machines~\cite{verl14,ging14,pole15,mart16}.

In this Letter, we find a new universal result about the statistics of
thermodynamic currents. We demonstrate that the fraction 
of time $\mathcal{T}_1$ that a generic thermodynamic
current stays above its average value (see Fig.~\ref{fig:1}) is distributed
according to Eq.~\eqref{eq:Levy}. This result is valid for 
mesoscopic systems  in a nonequilibrium steady state and also for  periodically-driven mesoscopic systems.
The proof of the arcsine law  for $\mathcal{T}_1$ is based on a theorem for Markov processes that has hitherto remained  
unexplored in physics~\cite{free63}. Our results are verified with experimental data from a Brownian 
Carnot engine~\cite{mart16}. Based on numerical evidence, we also conjecture that two other
random variables related to thermodynamic currents are distributed according to \eqref{eq:Levy}: the last time 
a fluctuating current crosses its average $\mathcal{T}_2$ and the time elapsed until a current reaches its maximal 
deviation from the average~$\mathcal{T}_3$.  
 


\underline{{\em Arcsine law for $\mathcal{T}_1$}}. We consider small nonequilibrium physical systems in contact 
with one or several thermal and/or particle reservoirs at thermal equilibrium. For instance, 
a single enzyme (the system) immersed in a solution
(the reservoir) that contains both substrate and product molecules. The system is in 
a nonequilibrium steady state if the concentrations of substrate and product in the large 
reservoir and the rate at which the enzyme consumes the substrate are approximately constant. 
In this example, the chemical potential difference between substrate and product is the 
thermodynamic force that drives the system out of equilibrium.

A vast class of these systems in physics and biochemistry  
can be described by Markov processes within the framework of stochastic thermodynamics~\cite{seif12}. 
In this framework,  thermodynamic currents take the form of integrated probability
currents. At steady state, their average rate is constant, leading to
a linear increase (or decrease) with time of the average thermodynamic currents. 
The fraction of time that a stochastic thermodynamic
current $X(t)$ spends above its average value $\langle X(t)\rangle$
during an observation time $t_f>0$ is defined as
\begin{equation}\label{eq:tt}
\mathcal{T}_1\equiv\frac{1}{t_f}\int_0^{t_f}  \theta\Big(X(t)-\langle X(t)\rangle\Big)\text{d}t,
\end{equation} 
where $\theta(x)$ is the Heaviside function. This random variable $\mathcal{T}_1$ is illustrated in Fig.~\ref{fig:1}.

\begin{figure}
\includegraphics[width=\linewidth]{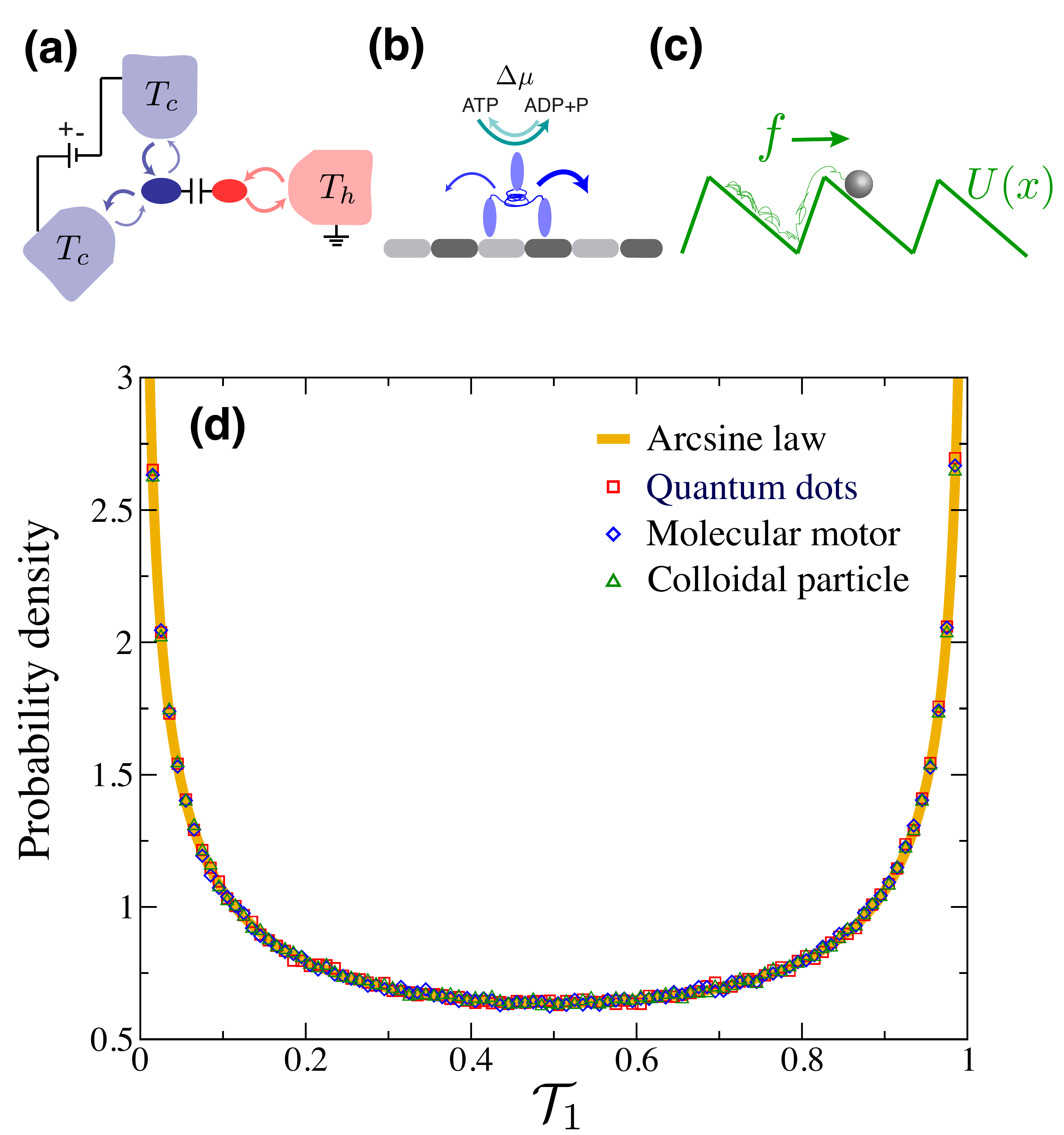}

\caption{{\bf Numerical illustration of the steady-state arcsine law for $\mathcal{T}_1$}. (a,b,c) Graphical illustration of the models.  (a) A double quantum dot.  (b) A molecular motor. (c) A colloidal particle on a periodic potential subjected to an external force. 
The thermodynamic currents that we consider are the electron current through the cold dot, the net number of steps of the motor,  and the net position of the particle, respectively. In (d) we compare the prediction of Eq.~\eqref{eq:Levy}  (orange line) with numerical simulations for
the three different models. For each curve, the number of realizations is $10^6$. Details of the three models are given in~\cite{supp}.
}
\label{fig:2} 
\end{figure}

Our main result is that for any thermodynamic current in a small system at steady state that is  described by a Markov process, the probability
density of $\mathcal{T}_1$, for large $t_f$, is given by
Eq.~\eqref{eq:Levy}. Hence, stochastic trajectories for which currents
such as heat, work, and entropy production stay all the time above or
below their average value are the most likely. The striking
universality of this result is illustrated in Fig.~\ref{fig:2}, where
we show numerical simulations of three models of different
physical systems: a double quantum dot~\cite{sanc13}, a molecular
motor~\cite{schm08a}, and a driven colloidal
particle~\cite{pigo17}. The mathematical proof of this result requires
the use of a theorem for Markov chains that establishes an arcsine law
for a random variable different from a current~\cite{free63}, and a suitable mapping between two Markov
chains~\cite{supp}. Interestingly, the proof also extends to time-symmetric  observables such as activity (or frenesy~\cite{baiesi2009fluctuations}) (see~\cite{supp} for details).


Small thermodynamic engines and several other systems of physical and
technological interest are driven by an external periodic protocol
\cite{mart17,erba15}. Such periodically-driven are described  by 
Markov process with time-periodic transition rates. Nevertheless, in the long time limit, it is possible to describe periodically-driven systems as steady
states of Markov processes with time-independent transition rates~\cite{bara16,ray17}. Hence, the
arcsine law for $\mathcal{T}_1$ is also valid for periodically-driven
systems, in the limit at which the observation time $t_f$ is much larger 
than the period of the protocol. We have illustrated this result with numerical simulations
of two models: a colloidal particle in a time-periodic potential and a
theoretical model for a Brownian Carnot engine~\cite{supp}.


\begin{figure}
\includegraphics[width=7.5cm]{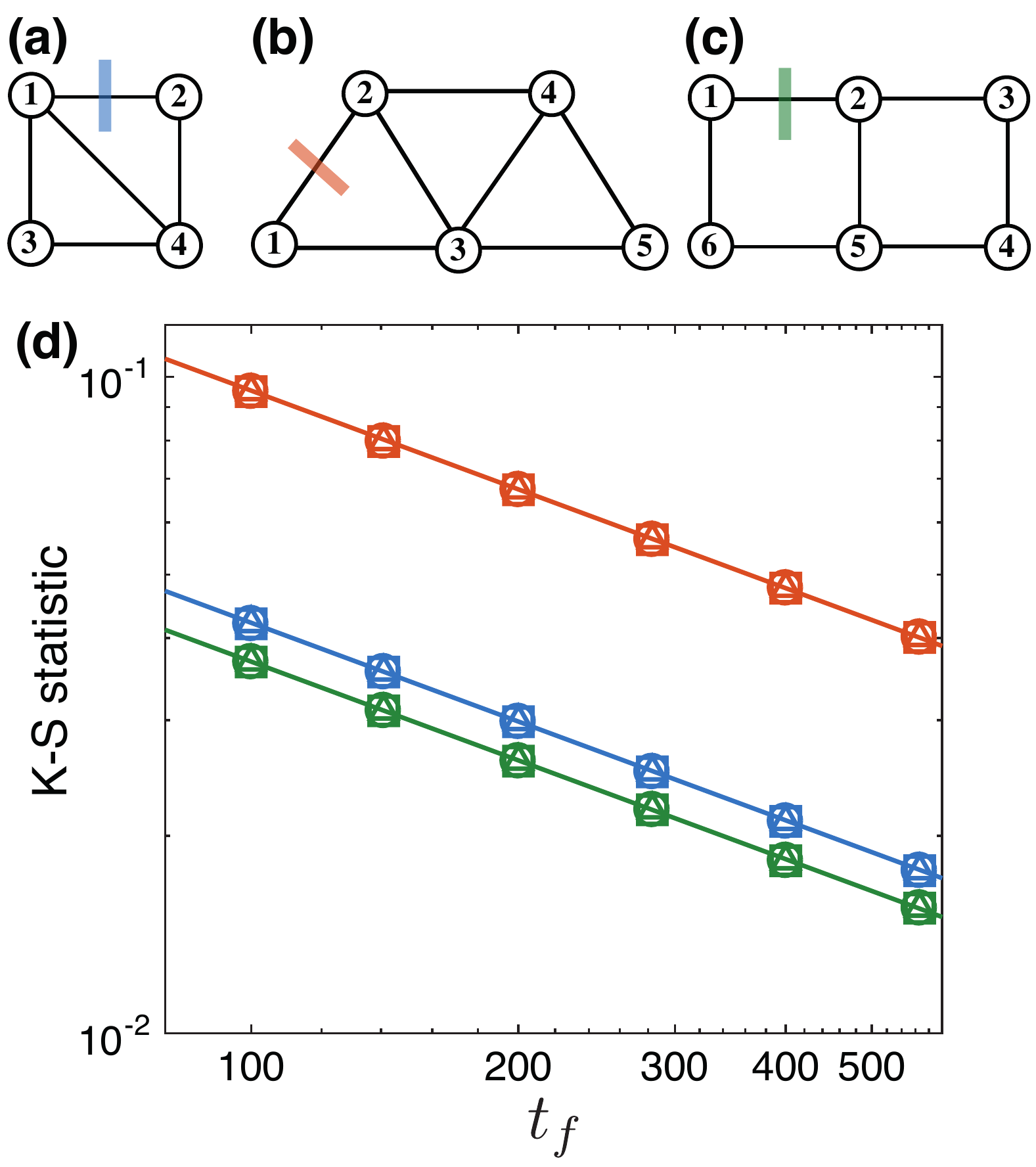}
\caption{\textbf{Numerical verification of the three arcsine laws.}   (a,b,c) Graphical illustration of minimal stochastic models with four (a), five (b) and six (c) different states. 
Each state is represented by a circle with a number and the lines represent non-zero transition rates. For all models we have evaluated the current from state 1 to state 2, as indicated with bars in the figures.
(d) Kolmogorov--Smirnov  statistic between a reference $\mathcal{T}$ described by the arcsine law~\eqref{eq:Levy} and $\mathcal{T}_1$ (triangles), $\mathcal{T}_2$ (circles) and $\mathcal{T}_3$ (squares), as a function of $t_f$. Different colors represent results for model a (blue), b (red) and c (green). Each symbol is obtained from $10^7$ numerical simulations of total duration $t_f$.   The lines are guides to the eye and are given by $\sim t_f^{-1/2}$.  Details of the three models are given in~\cite{supp}.
\label{fig:5}  }
\end{figure}

\underline{{\em Conjecture for $\mathcal{T}_2$ and  $\mathcal{T}_3$}}. For Brownian motion two other random variables obey Levy's arcsine law~\eqref{eq:Levy}. One is the last time the walker crosses zero and the other is the time the position of the walker reaches its maximum value. The equivalent random variables for the present case are defined as follows.
The fraction of time elapsed until a current crosses its average value for the last time $\mathcal{T}_2$ is  defined as 
\begin{equation}
\mathcal{T}_2\equiv \textrm{sup}_{t\in [0,t_f]}\left\{\frac{t}{t_f}\, : \, \Delta_X(t)=0\right\},
\label{eq:T2}
\end{equation}
where  $\Delta_X(t)\equiv X(t)-\langle X(t)\rangle$. The  time $t_{\rm sup}$ is defined as the time at which  $\Delta_X(t)$ attains its supremum, i.e., $\Delta_X(t_{\textrm{sup}})= \sup_{t\in [0,t_f]} \Delta_X(t)$. 
The fraction of time elapsed until a current reaches its maximal deviation above its average value is 
\begin{equation}
\mathcal{T}_3\equiv \frac{t_{\textrm{sup}}}{t_f}.
\label{eq:T3}
\end{equation}
We have verified numerically that both $\mathcal{T}_2$ and $\mathcal{T}_3$ are distributed according to~\eqref{eq:Levy}. Specifically, we have performed numerical simulations of the models shown in Fig.~\ref{fig:5}a-c with a finite observation time $t_f$, where $t_f$ is small enough such that we can accurately determine the third cumulant associated with the current, which is non-zero for all models (see \cite{supp}). Our simulations then probe large non-Gaussian fluctuations and, therefore, they test  arcsine laws 
for Markov processes, beyond Brownian motion. 

As shown in Fig.~\ref{fig:5}d, we have performed a finite-size scaling analysis of the K-S statistic for $\mathcal{T}_1$, 
$\mathcal{T}_2$, and $\mathcal{T}_3$, with respect to the arcsine distribution~\eqref{eq:Levy}, as a function of $t_f$. All random variables show the same behavior: for large times, the K-S statistic goes to zero as the power law $t_f ^{-1/2}$.
We then conjecture that $\mathcal{T}_2$ and $\mathcal{T}_3$ are also distributed according to~\eqref{eq:Levy}. 



\underline{{\em Experimental results}}. Heat engines are paradigmatic examples of periodically-driven systems~\cite{carnot1872reflexions}. 
We test the arcsine law for $\mathcal{T}_1$ using experimental data of a Brownian Carnot engine~\cite{mart16}. The working substance of the engine is a single optically-trapped colloidal particle of radius $R=500 \rm nm$ immersed in water. The particle is trapped in a time-periodic harmonic potential  $U(x, t) = \kappa(t) x^2(t ) /2$, whose stiffness $\kappa (t)$ is externally-controlled along a period $\tau$ between the minimum  value $\kappa_{\rm I}=\kappa(0)=(2.0\pm0.2){\rm pN\mu m^{-1}}$ and the maximum value $\kappa_{\rm III} =\kappa(\tau/2)=(20.0\pm0.2){\rm pN\mu m^{-1}}$. In addition, the kinetic temperature of the particle is switched periodically between a cold $T_{\rm c}=300\rm K$ and a hot temperature $T_{\rm h}=526 \rm K$. The temperature is controlled with an external noisy electrostatic field using the {\em whitenoise technique}~\cite{mart13}. The fine and simultaneous electronic control of the trap strength and the temperature of the particle allows us to implement protocols of different cycle times $\tau$ without loss of resolution, which range from $\tau=10\,\rm ms$ to $\tau=200\,\rm ms$. The total experimental time is $50 \rm s$ for all the values of $\tau$~\cite{supp}.

\begin{figure}
\includegraphics[width=8.5cm]{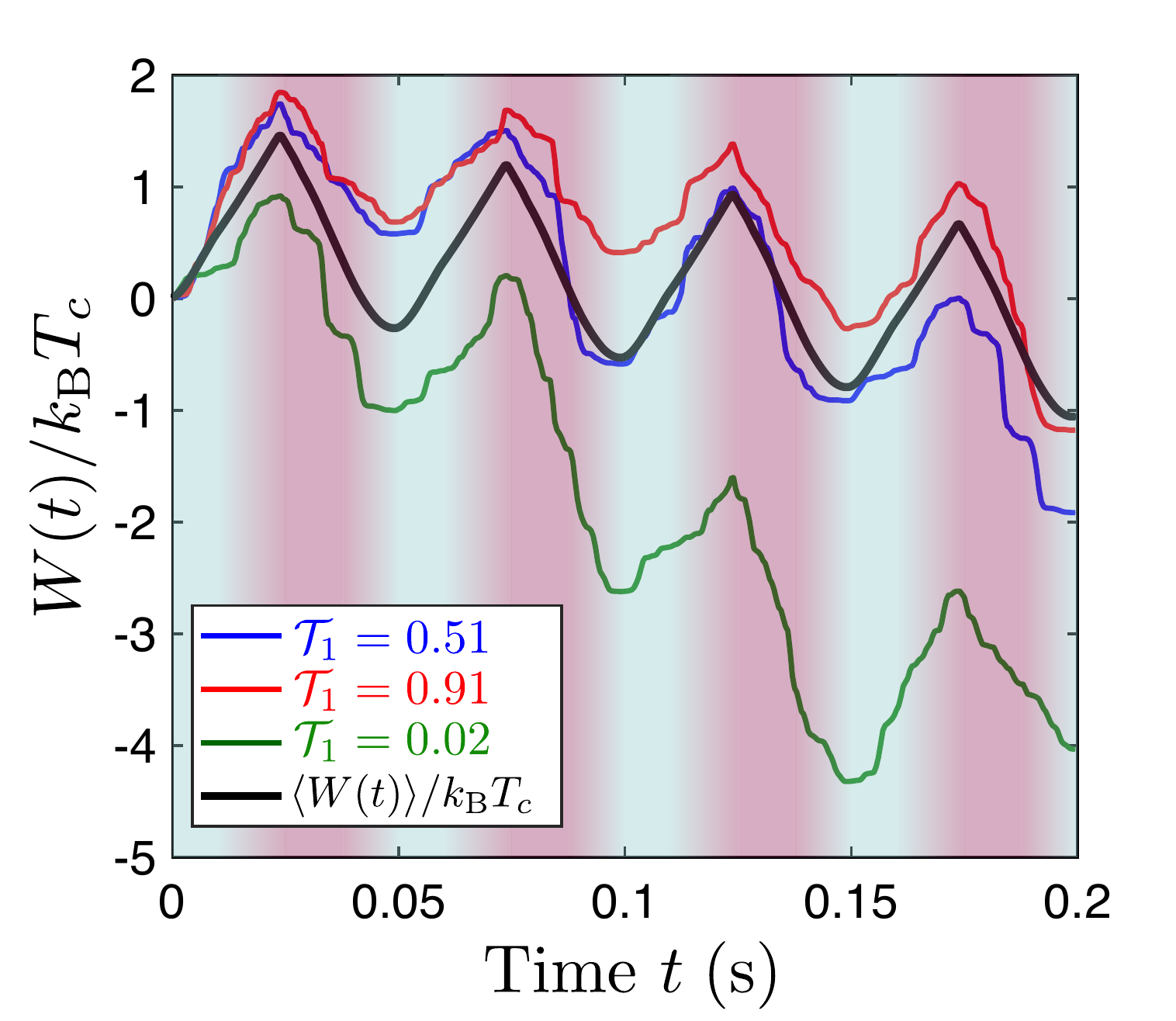}
\caption{\textbf{Fluctuations of  $\mathcal{T}_1$ in the Brownian Carnot engine.}   Sample traces of the stochastic work exerted on the colloidal particle as a function of time. The legend indicates the corresponding value of the time $\mathcal{T}_1$ elapsed for each trajectory above the average value of the work (black curve). The background color illustrates the temperature of the particle during the operation of the engine, with $T_c=300\,\rm K$ and $T_h=526\,\rm K$ corresponding to the minimum (blue) and maximum (red) temperatures of the engine. The isothermal steps are connected by microadiabatic protocols in which the temperature of the particle changes smoothly with time~\cite{mart15,supp}.
\label{fig:3} }
\end{figure}

\begin{figure}
\includegraphics[width=8.5cm]{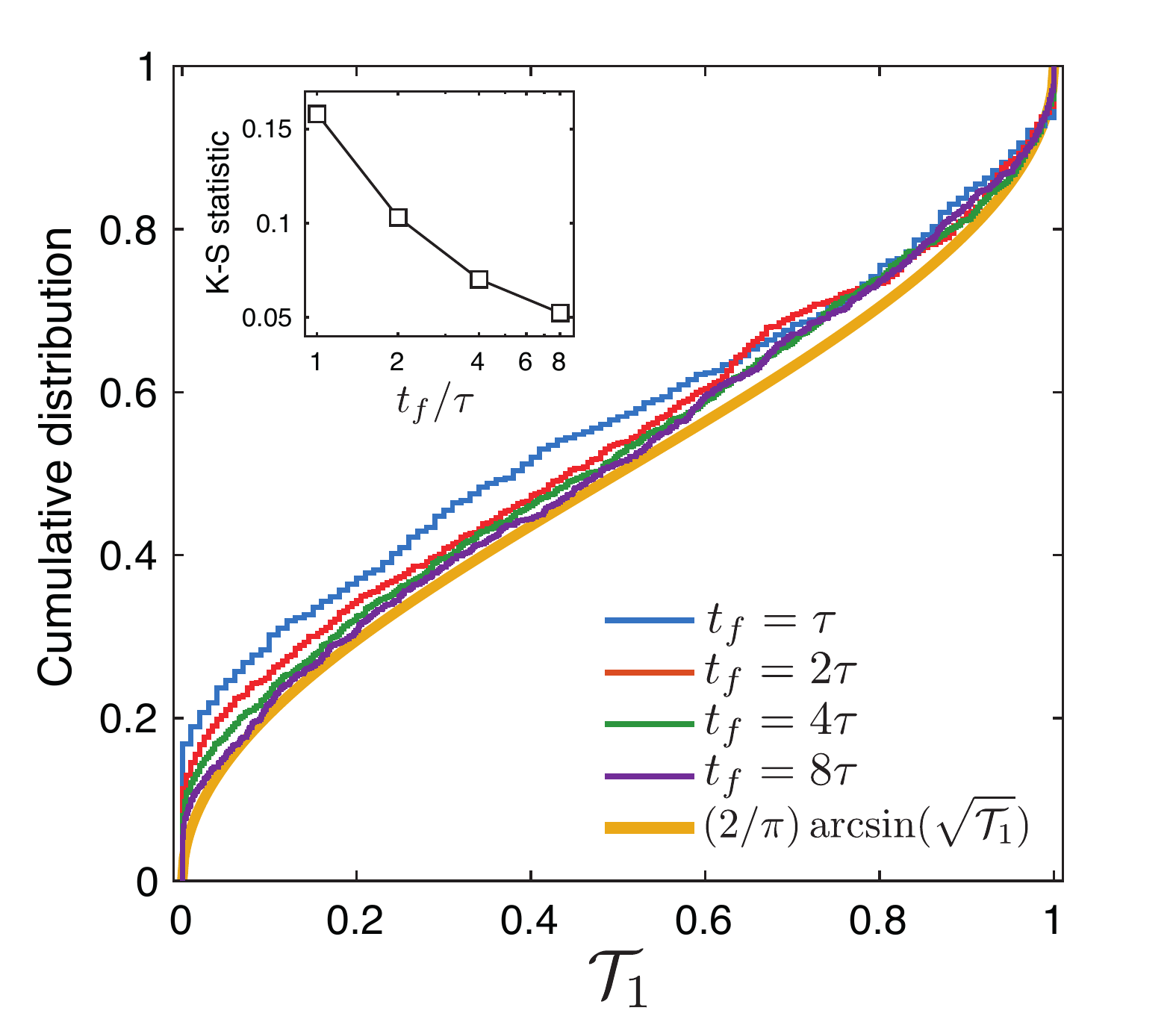}
\caption{\textbf{Experimental verification of the arcsine law for $\mathcal{T}_1$ in the Brownian Carnot engine.}  Empirical cumulative distribution of the fraction of time $\mathcal{T}_1$ the  work exerted on the particle elapses above its average value. The experimental data is obtained from $N=10^3$~cycles of the Brownian Carnot engine with cycle period  $\tau = 50\,\rm ms$. Different colors represent the experimental cumulative distribution of $\mathcal{T}_1$ calculated over different values of $t_f$ (see legend) and the theoretical arcsine distribution (thick orange line). Inset: Two-sample Kolmogorov-Smirnov (K-S) statistic~\cite{1933sulla,smirnoff1939ecarts} as a function of $t_f$. The line is a guide to the eye.
\label{fig:4}  }
\end{figure}

A key thermodynamic current that characterizes the performance of the Brownian Carnot engine is the stochastic work $W(t)$, where we adopt the convention that negative $W(t)$ means extracted work.
The stochastic work  is the change of $U(x(t))$ due to the external
control exerted on the particle that leds to a  time-varying stiffness
$\kappa(t)$ (see Eq.~\eqref{eq:Kt} in~\cite{supp}). We measure the work
from experimental traces of the particle position  by means of the expression 
 $W(t)=\int_0^t  (\partial U /\partial t') \text{d}t'=(1/2)\int_0^t  x^2(t')  \text{d}\kappa(t')$.

In order to test the arcsine law, we measure the
fluctuations of the fraction of time $\mathcal{T}_1$ that the stochastic
work $W(t)$ elapses above its average value, see Fig.~\ref{fig:3} for an
illustration.  We compute  $\mathcal{T}_1 = \frac{1}{t_f}\int_0^{t_f}
\text{d}t \,\theta(W(t)-\langle W(t)\rangle)$ integrating over
different values of the observation time $t_f$, which is an integer number of periods.  Since the arcsine law holds in the limit of $t_f$ large, we perform a finite-size-scaling analysis of the validity of Eq.~\eqref{eq:Levy}.  Figure~\ref{fig:4} shows that for the experimental data the cumulative distribution of $\mathcal{T}_1$  converges to $(2/\pi)\arcsin(\sqrt{\mathcal{T}_1})$ when increasing the observation time $t_f$. We quantify the discrepancies between the experimental data and numerical data generated with Eq.~\eqref{eq:Levy} using the two-sample Kolmogorov-Smirnov (K-S) statistic~\cite{1933sulla,smirnoff1939ecarts}. A finite-size-scaling analysis of the K-S statistic as a function of  $t_f$ reveals that the experimental distributions of $\mathcal{T}_1$ converge to the arcsine distribution. 
Notably, similar results are obtained   for different values of the period $\tau$ (see \cite{supp}).



\underline{{\em Conclusion}}. We have shown with theory, simulations and experiments that the fraction of time $\mathcal{T}_1$ a stochastic current elapses above (or below)
its average value is distributed according to Levy's arcsine law~\eqref{eq:Levy}. This result is valid for both systems in nonequilibrium
steady states and for periodically-driven systems such as mesoscopic engines.  
Based on numerical evidence, we have also conjectured that there are arcsine laws  for the last time $\mathcal{T}_2$ at which a current crosses its average value
and for the time $\mathcal{T}_3$ when a current reaches its maximal deviation from its average. 

 We have investigated fluctuations of mesoscopic systems described by Markovian dynamics. It is an open
question whether similar results also hold for non-Markovian stochastic processes used in the description of
active matter~\cite{bechinger2016active,prost2015active} and open quantum systems~\cite{breuer2016colloquium}. It will be interesting to investigate whether the arcsine laws for thermodynamic currents can be  used to design efficient control at the nanoscale.


 I.~A.~M. acknowledges financial support from Spanish Government, TerMic (FIS2014-52486-R) grant and Juan de la Cierva program. We acknowledge fruitful discussions with Izaak Neri, Raphael Chetrite and Hugo Touchette.

%
\bibliography{References}


\newpage\newpage\clearpage
\section{Supplemental Material}

This document provides additional information for the manuscript
``Arcsine Laws in Stochastic Thermodynamics''. It is
organized as follows. Section S1 contains the
derivation of the arcsine law for $\mathcal{T}_1$. Section S2 provides additional details on the stochastic models of molecular motor, quantum dot and colloidal particle used in Fig.~1 in the Main Text. Section S3 discusses the evaluation of the third cumulant for the currents in Fig.~5 in the Main Text.
Section S4 reports numerical simulations for periodically-driven systems. Section S5 provides further details of the experimental data. 

\section{S1. PROOF OF THE\\ FIRST  ARCSINE LAW}
\label{sec:proof}
In this section, we demonstrate the arcsine law for $\mathcal{T}_1$ using 
an arcsine law for Markov chains from~\cite{free63} together with a suitable
mapping. Let us consider an ergodic Markov chain defined by the discrete-time master equation
\begin{equation}
\Delta \mathsf{P}_i= \sum_{j}\left(\mathsf{P}_jT_{ji}-\mathsf{P}_iT_{ij}\right),
\label{eqmaster1}
\end{equation}    
where $i,j=1\dots\Omega$ are the states, $T_{ij}$ are the transition
probabilities from state $i$ to state $j$, $\Delta t$ is a discrete
timestep, $\mathsf{P}_{i}\equiv \mathsf{P}_{i}(n\Delta t)$ are the probabilities to be in state
$i$ at time $n \Delta t$, and $\Delta \mathsf{P}_i\equiv\mathsf{P}_{i}((n+1)\Delta t)-\mathsf{P}_{i}(n\Delta t)$.

An integrated fluctuating current is a functional of the stochastic trajectory 
$\{i_0,i_1\dots i_N\}$, where $N$ is the number of time steps, given by 
\begin{equation}
X_N\equiv\sum_{i,j} d_{ij}\mathcal{N}_{ij}.
\label{eqdb}
\end{equation}
where 
\begin{equation}
\mathcal{N}_{ij}\equiv \sum_{n=1}^{N} \delta_{i,i_{n-1}}\delta_{j,i_{n}}
\label{eqdb2}
\end{equation}
 counts the number of jumps from $i$ to $j$ in  $\{i_0,i_1\dots i_N\}$.
The increments $d_{ij}$ are anti-symmetric, i.e., $d_{ij}=-d_{ji}$. 
At steady state, the average time-integrated current is given by
 \begin{equation}
\langle X_N\rangle= N J\equiv N\sum_{i,j} d_{ij}\mathsf{P}_i^{\rm st}T_{ij},
\label{eq:dij}
\end{equation}
where  $\mathsf{P}_i^{\rm st}$ is the
stationary probability of state $i$. A physical interpretation of  $X_N$ as a thermodynamic current depends on 
the generalized detailed balance relation \cite{seif12}.

We now introduce an auxiliary Markov chain in the following way. Each
non-zero transition rate from state $i$ to state $j$ in the original
process is associated with a state $z_{ij}$ in the auxiliary Markov
process, so that the number of states in the auxiliary process is $Z\le\Omega^2$.
The master equation for this auxiliary process reads 
\begin{equation}
\Delta \mathsf{P}(z_{ij})= T_{ij}\sum_k\mathsf{P}(z_{ki})-\mathsf{P}(z_{ij})\sum_kT_{jk},
\label{eqauxil}
\end{equation}   
where $\mathsf{P}(z_{ki})$ is the probability for the auxiliary process to be in state $z_{ij}$ at time $n\Delta t$. 
There is a bijective relation between trajectories
$\{i_0,i_1\dots i_N\}$ of the original process and
trajectories $\{z_{i_0 i_1},z_{i_1 i_2},\dots z_{i_{N-1} i_N}\}$ of 
the auxiliary process. In particular, the number of transitions $\mathcal{N}_{ij}$ in Eq.~\eqref{eqdb2} for
the original process becomes the number of time steps that a state $z_{ij}$ is visited in the auxiliary 
process, i.e.,
\begin{equation}
\mathcal{N}_{ij}=\sum_{n=1}^{N} \delta_{z_{ij},z_{i_{n-1}i_{n}}}.
\label{eqdb3}
\end{equation}
Therefore, the current~\eqref{eq:dij} can be written as
\begin{equation}
X_N\equiv\sum_{z_{ij}} f(z_{ij})\mathcal{N}_{ij},
\end{equation}
where $f(z_{ij})\equiv d_{ij}$ is a real function on the space of states. 
Its steady-state average~\eqref{eq:dij} then becomes 
\begin{equation}
\langle X_N\rangle = N \sum_{z_{ij}} f(z_{ij})\mathsf{P}^{\rm st}(z_{ij}).
\end{equation}
Comparing the average over all trajectories of $\mathcal{N}_{ij}$ for the auxiliary process~\eqref{eqdb3} with the same average for the original process~\eqref{eqdb2} we find
\begin{equation}
\mathsf{P}(z_{ij})= \mathsf{P}_i T_{ij}.
\end{equation}
Note that Eq.~\eqref{eqmaster1} can be obtained from 
Eq.~\eqref{eqauxil} by means of the relation $\mathsf{P}_i=\sum_k\mathsf{P}(z_{ki})$.

We now prove our main result with the auxiliary process. 
We define a functional $\theta_n$, which indicates whether $X_n>\langle X_n \rangle$, with the help of  the sum over a trajectory  
\begin{align}
S_N &\equiv  \sum_{z_{ij}} f(z_{ij})\mathcal{N}_{ij} -\sum_{n=1}^N \langle X_n\rangle \nonumber\\
&= \sum_{n=1}^N \left(f(z_{i_{n-1} i_n}) - \langle X_n\rangle\right),
\end{align}
as follows
\begin{equation}
\theta_n = \begin{cases} 0 &\mbox{if } S_n \leq 0 \\ 
1 & \mbox{if } S_n>1 \end{cases} .
\end{equation}
 The random variable 
\begin{equation}
\widehat{\mathcal{T}}_N\equiv N^{-1} \sum_{n=1}^N \theta_n
\end{equation}
counts the fraction of the $N$ time steps for which  $X_n>\langle X_n\rangle$.
From Theorem 1 of \cite{free63}, 
 the probability density associated with $\widehat{\mathcal{T}}_N$ in the limit $N\to \infty$ is 
given by Eq.~\eqref{eq:Levy}. Two key assumptions for this theorem are $\langle {S}_N\rangle=0$ and $\lim_{N\rightarrow\infty}\text{Prob}({S}_N>0)=1/2$, which are a consequence of central limit theorem for Markov chains~\cite{jone04}.  

In the above demonstration there is no need to assume an antisymmetric $d_{ij}$. For instance, the arcsine law is also valid for quantities like activity (or frenesy~\cite{baiesi2009fluctuations})
that count number of transitions between states, which corresponds to a symmetric $d_{ij}$.

The arcsine law should also hold for Langevin equations and for periodically-driven systems due to 
the following  arguments.
First, overdamped Langevin equations can be obtained as a limit of a master
equation with a large number of states, hence the arcsine law should also hold.  
Second, periodically-driven systems 
with stochastic protocols can be analyzed within the steady state of a bipartite Markov process \cite{bara16,ray17}. 
Since the proof above is also valid for such bipartite Markov processes, and, in the long time limit, a deterministic protocol can be obtained as 
limit of a stochastic protocol with many jumps, we have a justification for the arcsine law for periodically-driven systems.

The models we used in our numerical simulations are continuous-time models. In this case,  
$\Delta t\to 0$, which leads to transition rates $w_{ij}=\lim_{\Delta t\to 0} T_{ij}/\Delta t$, and  observation time 
$t_f=N\Delta t$. For models with a discrete state space, the large time limit for which the arcsine law holds corresponds to an observation time $t_f$ much larger than the maximum inverse escape rate in the system. For periodically-driven systems, $t_f$ must be much larger than the period $\tau$.  


\section*{S2. STOCHASTIC MODELS OF DOUBLE QUANTUM DOT, MOLECULAR MOTOR AND DRIVEN COLLOIDAL PARTICLE}

In this section we describe the models used in the numerical simulations shown in Fig.~2 of the Main Text.  The models for a
molecular motor and a double quantum dot are described by a continuous-time
master equation.

\begin{figure}
\includegraphics[width=80mm]{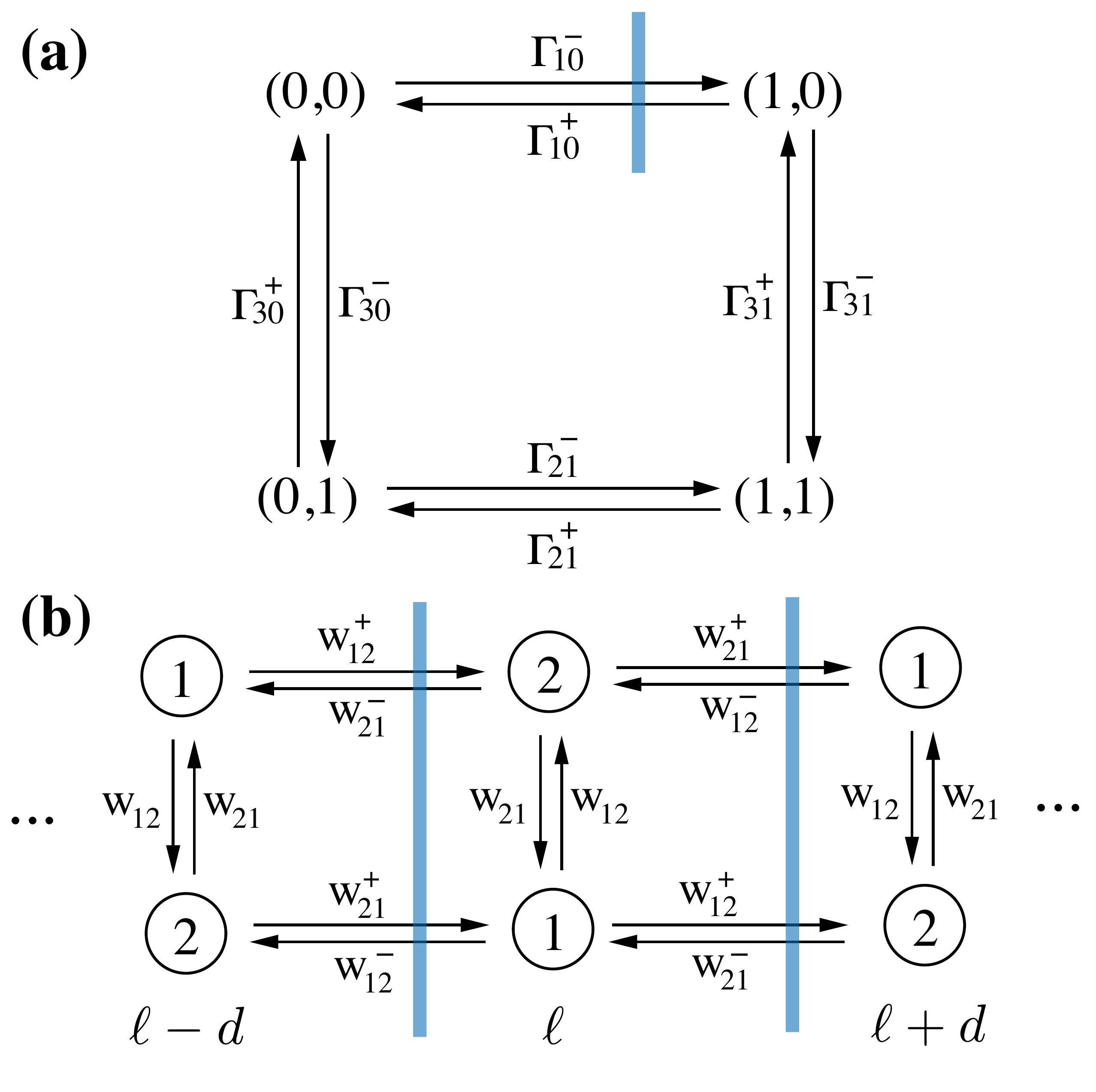}
\caption{\textbf{Illustration of the Markov models for (a) a double quantum dot and (b) a molecular motor.} For the double quantum dot the current is the net number of jumps between (0,0) and (1,0) and for the molecular motor the net number of jumps from left to right as indicated in the figure. For both cases the current is given by the net number of jumps crossing the blue line from left to right.
\label{figasinlawsmodels} }
\end{figure}

The basic physics of the double quantum dot model~\cite{sanc13} is
the following. We consider two quantum dots that can be either
occupied by an extra electron or empty, leading to four states
$(n_s,n_g)$, with $n_{s,g}=\{0,1\}$. Electrons can tunnel between each dot and
the external reservoir.  The first quantum dot $s$ is connected to two
reservoirs at an inverse temperature $\beta_c$. The voltages of
reservoirs $1$ and $2$ are $V_1$ and $V_2$, respectively. The second
quantum dot $g$, which is capacitively coupled to the dot $s$, is
connected to a third reservoir at an inverse temperature
$\beta_h\le \beta_c$ and a voltage $V_3$.  The energy of an occupied
dot reads $E_n= E_0+nE_{\textrm{int}}$, where $n=0$ ($n=1$) if the
other dot is empty (occupied) and $E_{\textrm{int}}$ is an interaction
energy. The interaction between the two dots generates correlations
that can lead to thermoelectric transport through dot $s$ against the
voltage difference $V_2-V_1$. We consider a regime named ``optimal
configuration'' in \cite{sanc13}. In this regime there is tight
coupling between the heat flux and the electron flux.

\begin{figure*}
\centering
\includegraphics[width=180mm]{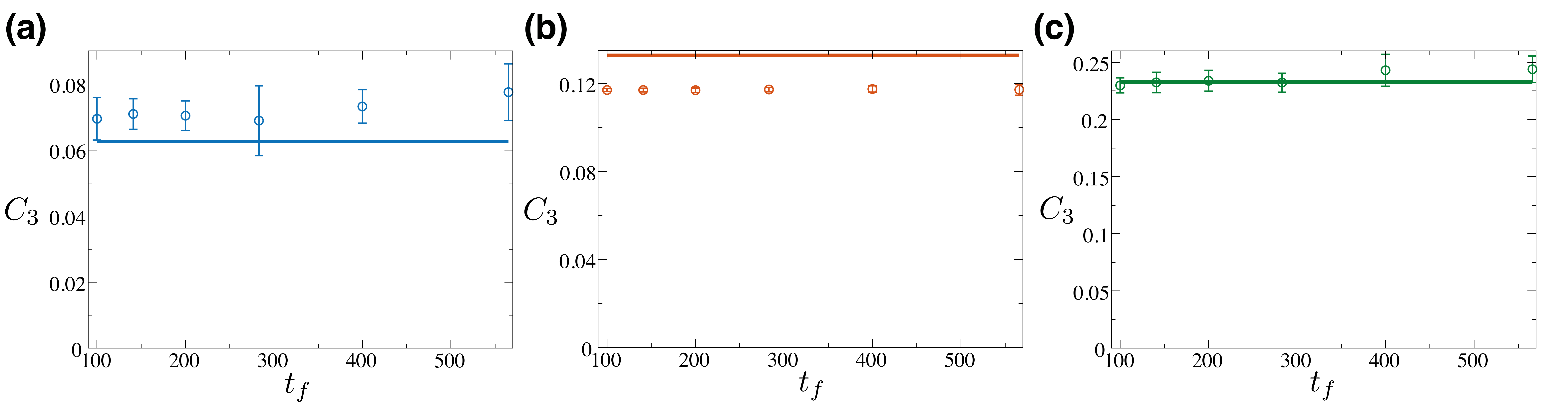}
\caption{\textbf{Numerical evaluation of third cumulant for the models depicted in  Fig.~5a-c in the Main Text.}
(a) 4-state model. (b) 5-state model. (c) 6-state model. The value of the rates for our
simulations are given in the caption of Fig.~5 in the Main Text. The solid lines indicate the exact value of C3 for $t_f\to\infty$~\label{fig:c3}. }
\end{figure*}

The transition rates $\Gamma^{\pm}_{\alpha n}$ for this model are illustrated in
Fig.~\ref{figasinlawsmodels}.  The first subscript in the tunnelling
rates $\alpha=1,2,3$ refers to the reservoir and the second subscript
 $n=0,1$ refers to whether the other dot is empty or occupied,
respectively. The superscript $-$($+$) denotes an electron tunnel
into (out of) the dot. The tunneling rates are
$\Gamma^-_{\alpha n}= f(\beta_\alpha(E_n-V_\alpha))$ and
$\Gamma^+_{\alpha n}=1-f(\beta_\alpha(E_n-V_\alpha))$, where
$f(x)=(1+\textrm{e}^x)^{-1}$ is the Fermi function. We consider as a
thermodynamic current the number of transitions from state $(0,0)$ to
state $(1,0)$ minus the number of transitions from from state $(1,0)$
to state $(0,0)$. This current is proportional to both the heat and
electron flux. For the results shown in Fig. \ref{fig:2} the
parameters are $\beta_h=1$, $\beta_c=4.63$, $E_0= 0.05$,
$E_{\textrm{int}}=1$, $V_1= 0.525$, $V_2=-0.1375$, and $V_3=0$.

We next consider a model of a chemically-driven molecular motor in the
presence of an external force~\cite{schm08a}. Part of the chemical
work obtained from ATP hydrolysis drives the motor against the 
mechanical force. The state of the motor is specified by its position
$\ell$ and its conformational state that can be either $1$ or $2$. The
energy difference between the conformational states is $\Delta E$, where the
inverse temperature is set to $\beta=1$.  The free energy of one ATP
hydrolysis is $\Delta \mu$, the mechanical force is $F$ and the motor
stepsize is $d$. The transition rates are represented in
Fig. \ref{figasinlawsmodels}.  Rates
of conformational change that involve ATP hydrolysis are
$w_{12}=w\textrm{e}^{(\Delta\mu-\Delta E)/2}$ and
$w_{21}=w\textrm{e}^{-(\Delta\mu-\Delta E)/2}$. The transition rates
for a forward step $\ell\to \ell+d$ are
$w_{12}^+=k_1\textrm{e}^{-\Delta E/2-Fd}$ and
$w_{21}^+=k_2\textrm{e}^{\Delta E/2-Fd}$. Finally, the transition
rates for a backward step $\ell\to \ell-d$ are
$w_{12}^-=k_2\textrm{e}^{-\Delta E/2}$ and
$w_{21}^-=k_1\textrm{e}^{\Delta E/2}$. The current we choose is the
position of the motor. Whenever a jump associated with a transition
rate with the superscript $+$ ($-$) occurs, this current increases
(decreases) by one. For the results shown in Fig. \ref{fig:2}, the
parameters are $\Delta E=2$, $\Delta \mu=20$, $Fd=1$, $w=10$, $k_1=1$,
and $k_2=1/2$.

The model for a colloidal particle in a periodic potential is
described by the overdamped Langevin equation

\begin{equation}
\frac{\text{d}x(t)}{\text{d}t}=\mu \left[f-\partial_x U(x(t))\right]+\sqrt{2D}\xi(t)
\end{equation}
where $\mu$ is the mobility, $f$ is a costant external force, $U(x)$ a
periodic energy potential, $D$ the diffusion coefficient and $\xi(t)$ a delta-correlated noise source with $\langle \xi(t)\rangle=0$ and $\langle \xi(t)\xi(t')\rangle=\delta (t-t')$.
We consider the periodic potential $U(x)=x/x^*$ for $x\le x^*$ modulo
$1$ and  $U(x)=(1-x)/(1-x^*)$ for $x> x^*$ modulo
$1$, with $x^*=1/3$. The other parameters that we used in Fig.~2 in the Main Text 
are $\mu=D=1$ and $f=2$.

\section{S3. EVALUATION OF THIRD CUMULANT FOR FIG. 5}

We have evaluated with numerical simulations the third cumulant associated with the currents $X$ indicated in Fig. 5 in the Main Text. Since, this quantity is extensive with
the observation time $t_f$, we have calculated
\begin{equation}
C_3\equiv \left\langle (X_f-\langle X_f \rangle)^3\right\rangle/t_f,
\end{equation}
where $X_f\equiv X(t_f)$ and the brackets indicate an average over stochastic trajectories.
For all the models shown in Fig.~5 in the Main Text, we have performed $10$ independent numerical evaluations of $C_3$, where each contains $10^7$ realizations.
The results are shown in Fig.~\ref{fig:c3}. The error bars are the mean standard deviation calculated with the $10$ independent simulations.
Clearly, $C_3$ is non-zero for all observation times $t_f$ used for the finite-size scaling in the main text. Hence,
our simulations do probe large deviations beyond the Gaussian regime. We have also compared the numerical results with the exact
value of $C_3$, which can be evaluated from the maximum eigenvalue of a modified generator~\cite{lebo99}. Even though the numerical
results are compatible with the exact result in Fig.~\ref{fig:c3}, there is an apparent systematic discrepancy, which is due to the
finite observation times $t_f$. 

The parameters for Fig.~5 in the Main Text and Fig.~\ref{fig:c3} were set to: for model a, $w_{12}= 2.5$, $w_{13}= 3.0$, $w_{14}= 0.33$,
$w_{21}= 1.7$, $w_{24}= 2.8$, $w_{31}= 2.0$, $w_{34}= 5.1$, $w_{43}= 4.7$, $w_{42}= 5.2$, and $w_{41}= 5$;
for model b, $w_{12}= 10.5$, $w_{13}= 3.2$, $w_{21}= 0.54$, $w_{23}= 1.2$, $w_{24}= 4.8$, $w_{31}= 2.5$, $w_{32}= 5.7$ $w_{34}= 14.7$,
$w_{35}= 31$, $w_{42}= 5.2$, $w_{43}= 2.7$, $w_{45}= 7.55$, $w_{53}= 27$, and $w_{54}= 40$; for model c,
$w_{12}= 50.5$, $w_{16}= 3.2$, $w_{21}= 1.54$, $w_{23}= 0.28$, $w_{25}= 4.8$, $w_{32}= 5.7$, $w_{34}= 2.5$, $w_{43}= 14.7$, $w_{45}= 31$, $w_{52}= 5.2$, $w_{54}= 2.7$,
$w_{56}= 7.55$, $w_{61}= 27$, and $w_{65}= 40$.

\section{S4. ARCSINE LAW FOR PERIODICALLY-DRIVEN SYSTEMS: NUMERICAL SIMULATIONS}

We now report on numerical simulations for periodically-driven systems.
From the mathematical argument
presented in Sec.~S1 the arcsine law should
also apply to these systems, provided that the observation time $t_f$ is large compared with
the period of the driving protocol. We illustrate this idea in a
simple model of a periodically-driven colloidal system, and then study
a more complex model which describes our experimental setup

\begin{figure}
\includegraphics[width=80mm]{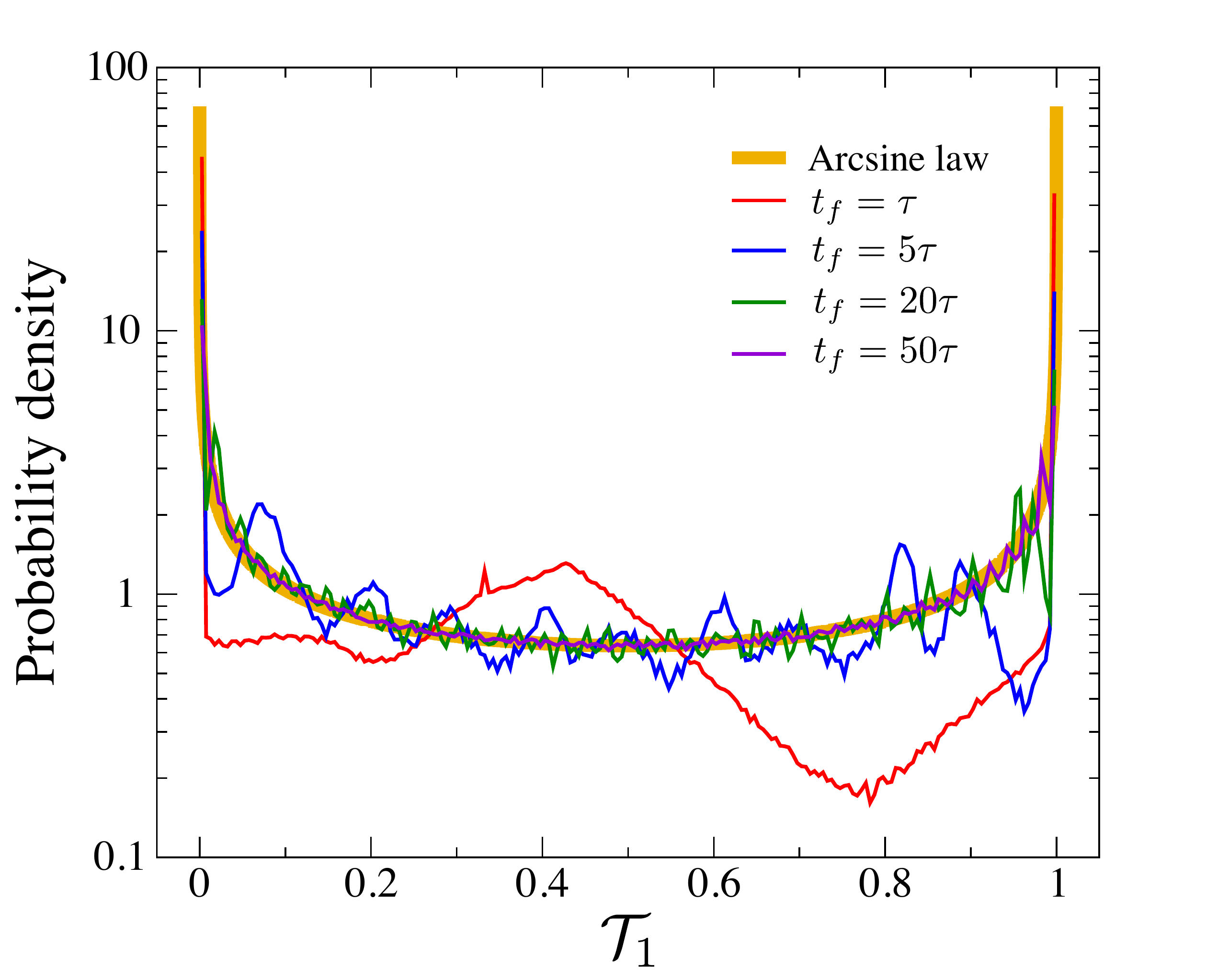}
\caption{\textbf{Numerical verification of the arcsine
    law for $\mathcal{T}_1$ for a periodically-driven Langevin system.} The dynamics of the system
  is given by Eq.~\eqref{eq:simpleperiodic},  with
  $\mu=\kappa_0=\nu=1$.  Different curves show the distribution of $\mathcal{T}_1$ for the work 
  $W(t)=\int_0^t (\partial U/\partial \kappa)\text{d}\kappa(t')$ to be above its average value 
  for different values of $t_f$ (see legend). 
  Notice the convergence to the arcsine law (orange line) as $t_f$ is increased.\label{periodicsimple}}
\end{figure}

We first consider a colloidal particle confined in a
harmonic trap $U(x,t)=(1/2)\kappa(t)x^2$ with periodically varying stiffness $\kappa(t)=\kappa_0 [2+\sin(2\pi \nu t)]$ described by the overdamped Langevin equation
\begin{eqnarray}\label{eq:simpleperiodic}
\frac{\text{d}x(t)}{\text{d}t}&=&-\mu k(t) x(t)+\sqrt{2D}\xi(t),
\end{eqnarray}
where $\xi(t)$ is a delta-correlated Gaussian white noise of zero mean and amplitude one.
The Langevin equation~\eqref{eq:simpleperiodic} is numerically integrated (using Euler's numerical integration scheme). After a periodic
steady-state has been reached, the statistics  of 
$\mathcal{T}_1$, associated with the work exerted to the particle defined in the caption of Fig.~\ref{periodicsimple}, are computed over different observation times $t_f$ which are integer multiples of the period $\tau$. Figure~\ref{periodicsimple} shows that for increasing $t_f$ the distribution of $\mathcal{T}_1$ converges to Eq.~\eqref{eq:Levy}.

\begin{figure}
\includegraphics[width=75mm]{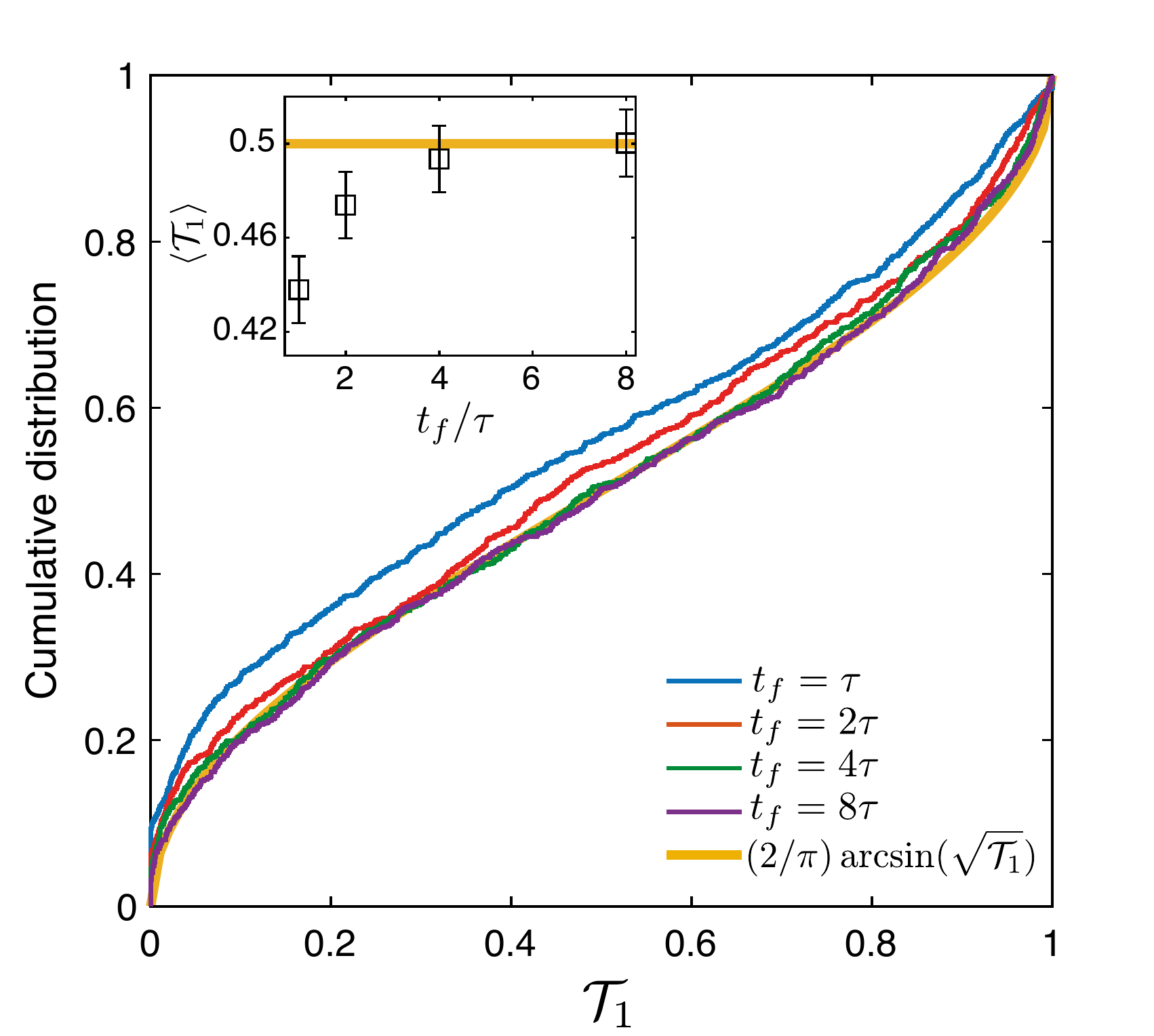}
\caption{\textbf{Numerical verification of the arcsine law for $\mathcal{T}_1$ for a periodically-driven Brownian Carnot engine.} Empirical cumulative distribution of the time $\mathcal{T}_1$ that the stochastic work elapses above its average obtained from numerical simulations of the Brownian Carnot engine described by Eqs.~(\ref{eq:le}-\ref{eq:Tt}). Distributions for the fraction of time $\mathcal{T}_1$ over a total observation time $t_f$ corresponding to $1$, $2$, $4$ and $8$ cycles of the engine are compared with the theoretical arcsine cumulative distribution (see legend). Inset: Value of the mean value of $\mathcal{T}_1$ as a function of the observation time. The orange horizontal line is set at the theoretical value  $\langle\mathcal{T}_1\rangle=1/2$ corresponding to the mean value of~Eq.~\eqref{eq:Levy}. Parameters of the simulations: mass $m=0$, friction coefficient $\gamma =8.4\rm\,pN ms/\mu m$, simulation time step $\Delta t = 1\mu s$, sampling frequency $f=2\,\rm kHz$,  period $\tau = 100\,\rm ms$ and total time $t_{\rm sim} = 100\,\rm s$.
\label{fig:carnotsim} }
\end{figure}

We next consider a minimal model for the Brownian Carnot engine~\cite{mart15} for which we report experimental data. The model is given by the following Langevin equation
\begin{equation}
m\ddot{x}(t) = - \gamma \dot{x}(t) - \kappa(t) x(t) + \sqrt{2k_{\rm B}T(t) \gamma}\xi(t),
\label{eq:le}
\end{equation}
which  models the dynamics of a Brownian particle with mass $m$, immersed in a thermal bath with friction $\gamma$. The particle is trapped with a harmonic potential of time-periodic strength $\kappa(t)$ and immersed in a thermal bath of time-periodic temperature $T(t)$~\cite{mart15,mart16}. 
 The trap stiffness is modulated in time following a time-symmetric  discontinuous protocol of period $\tau$
\begin{eqnarray}
\kappa(t) & =\left\{
\begin{array}{ll}
\kappa_0 + \alpha t^2 &\; \text{for} \;\;t\in[0,\tau/2)\,\\
\kappa_0 + \alpha (t-\tau)^2& \; \text{for} \;\;t\in[\tau/2,\tau)\,,\\
               \end{array}
        \right.    \label{eq:Kt}
\end{eqnarray} 
where $\kappa_0=2\,\rm pN/\mu m$ is the initial trap stiffness and $\alpha = 4(\kappa_2-\kappa_0)/\tau^2$, with  $\kappa_2=20\,\rm pN/\mu m$. 
The temperature  follows a time-asymmetric protocol of period $\tau$ given by
\begin{eqnarray}
T(t) & =\left\{
\begin{array}{ll}
T_c &\; \text{for} \;\;t\in[0,\tau/4)\,\\
\displaystyle T_c\,\sqrt{\frac{\kappa_0 + \alpha t^2}{\kappa_1}} &\; \text{for} \;\;t\in[\tau/4,\tau/2)\,\\
T_h &\; \text{for} \;\;t\in[\tau/2,\tau^{\star})\,\\
\displaystyle T_c\,\sqrt{\frac{\kappa_0 + \alpha (t-\tau)^2}{\kappa_0}} & \; \text{for} \;\;t\in[\tau^{\star},\tau)\,,\\
               \end{array}
        \right.    \label{eq:Tt}
\end{eqnarray} 
where $T_c=300\,\rm K$, $T_h=T_c \sqrt{\kappa_2/\kappa_0} = 526.3\,\rm K$, and $\tau^{\star}\simeq 0.76\tau$. The first step in~\eqref{eq:Tt} corresponds to an cold isothermal compression, the second step to a {\em microadiabatic}~\cite{mart15} compression, the third step corresponds to a hot isothermal expansion and the fourth step to a microadiabatic expansion.

We perform numerical simulations of Eq.~\eqref{eq:le} under the periodic driving of both trap strength~\eqref{eq:Kt} and temperature modulation~\eqref{eq:Tt}. In our simulations, we measure the fraction of time $\mathcal{T}_1 = \frac{1}{t_f}\int_0^{t_f} \text{d}t \theta(W(t)-\langle W(t)\rangle)$ that the work exerted on the particle  $W(t)=\int_0^t  (\partial U /\partial t') \text{d}t'=(1/2)\int_0^t  x^2(t')  \text{d}\kappa(t')$ is above its average value $\langle W(t)\rangle$. 
Our numerical results show that the cumulative distribution of $\mathcal{T}_1$ tends to 
the cumulative distribution $F(\mathcal{T}_1)=(2/\pi)\arcsin(\sqrt{\mathcal{T}_1})$ (see Fig.~\ref{fig:carnotsim}), when increasing the observation time $t_f$, 
in agreement with the arcsine law for $\mathcal{T}_1$. The inset of Fig.~\ref{fig:carnotsim} shows that the mean value of $\mathcal T_1$ converges to $1/2$, in agreement with the average of the distribution given by Eq.~(1) in the Main Text.

\section*{S5. EXPERIMENTAL DATA}

\begin{figure}[ht!]
\includegraphics[width=70mm]{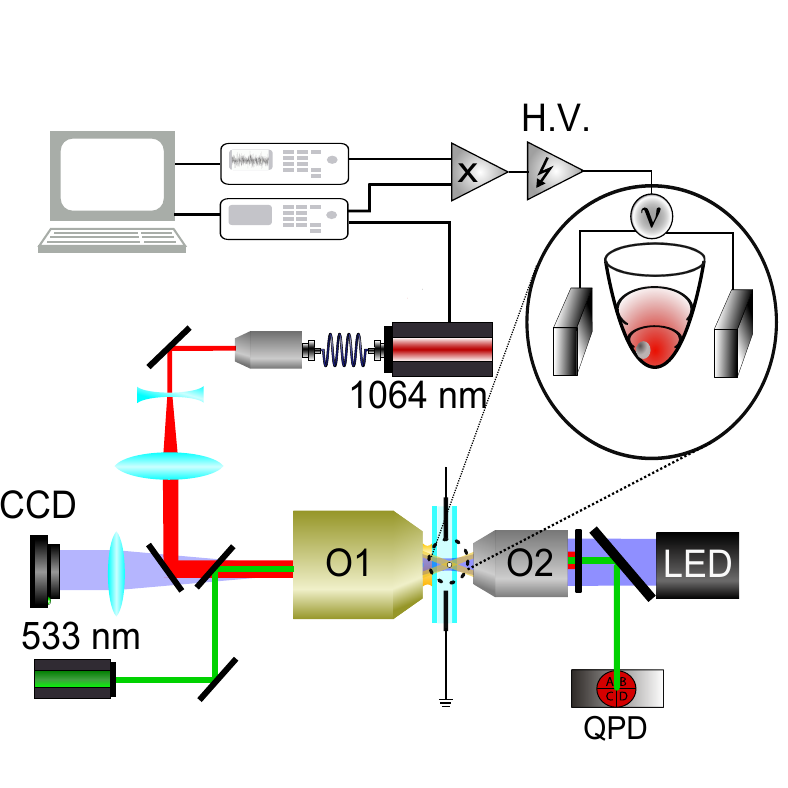}
\caption{\textbf{Sketch of the experimental setup.} A single microscopic colloid (grey sphere) is trapped with an optical-tweezer setup constructed with an infrared trapping laser (red). The position of the particle is tracked with an additional detection laser (green). An external noisy electrostatic field controls the amplitude of the fluctuations of the particle following the whitenoise technique~\cite{mart13}.
\label{fig:experiment} }
\end{figure}

Figure~\ref{fig:experiment} illustrates the experimental setup used to test the arcsine law for $\mathcal{T}_1$ which was previously described~\cite{mart16}. The setup is based on a horizontal self-built inverted microscope, where the sample is illuminated by a white lamp while the image is captured by a CCD camera. An infrared diode laser (wavelength $\lambda=980\,\rm nm$)  is highly focused by a high numerical aperture (NA) immersion oil objective to create the optical trap. A laser controller (Arroyo Instruments 4210) controls the optical power, and therefore the trap strength $\kappa$, at a maximum rate of $250\,\rm kHz$ using an external voltage $V_{\kappa}$. 

Polystyrene beads (G. Kisker-Products for Biotechnology, PPs-1.0, diameter $d=(1.00\pm0.05)\,\rm\mu m$) are diluted in Milli-Q water to a final concentration of a few microspheres per mL. The solution is injected into a custom-made electrophoretic chamber. Two aluminium electrodes are placed at the two ends of the chamber to apply a controllable voltage to the sample. The applied voltage is a computer-generated Gaussian white noise signal of amplitude $V_T$~\cite{mart13}.
Both $V_{\kappa}$ and $V_T$ are controlled by the same signal generator (Tabor electronics, WW5062) run by a custom-made LabView software. In the case of $V_T$, the output signal of the signal generator is amplified $10 00$ times with a high-voltage power amplifier (TREK, 623B). 

The particle is tracked using an additional green laser (wavelength $\lambda=532\,\rm nm$) collimated by a microscope objective ($\times$10, NA 0.10) and sent through the trapping objective O1. The light scattered by the trapped object is collected by the objective O2 (Olympus, 40$\times$, NA 0.75) and projected into a quadrant photo detector (QPD, Newfocus 2911), which has maximum acquisition frequency  of $200\,\rm kHz$. The signal is transferred through an analog-to-digital conversion card (National Instruments PCI-6120).

The nano-detection system is calibrated using the statistics of the thermal fluctuations of the bead trapped with a static trap at room temperature~\cite{visscher1996construction}. The input voltage controls the noise intensity and can be related to the effective temperature of the particle $T=\kappa \langle x^2\rangle / k_{\rm B}$ as $T=T_{\rm water}+S_T V^2_T$, where $S_{T}$(K/V$^2$) is the calibration factor and $T_{\rm water} = 300\,\rm K$ the temperature of the water. 

Using the fine simultaneous control of  $V_{\kappa}$ and $V_T$ we implement thermodynamic cycles of periods ranging from $\tau=20\,\rm ms$ to $\tau=200\,\rm ms$ during a total experimental time $t_{\rm exp}=50\,\rm s$ following the time-dependent protocols for the trap stiffness $\kappa(t)$ and bead temperature $T(t)$ given by Eqs.~\eqref{eq:Kt} and~\eqref{eq:Tt}, respectively.  Traces of the bead position $x(t)$ with sampling frequency $f_{\rm acq}=2\,\rm kHz$ are obtained using the aforementioned calibration methods and used to calculate the statistics of the work done on the bead.  

\begin{figure*}
\includegraphics[width=\textwidth]{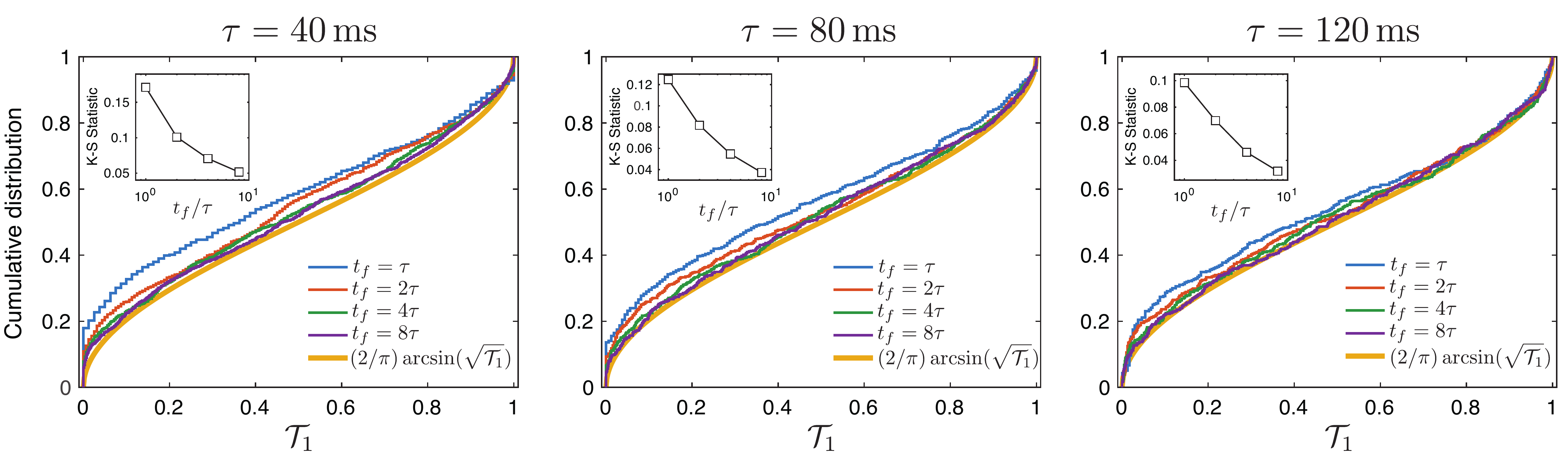}
\caption{\textbf{Experimental test of the arcsine law for $\mathcal{T}_1$ in the Brownian Carnot engine for different values of the cycle period $\tau$.}  Cumulative distribution of the fraction of time $\mathcal{T}_1$ the stochastic work  elapses above its average value calculated for different values of the observation time $t_f$ (blue, red, green, purple, see legend) compared to the theoretical prediction from the arcsine law $F(\mathcal{T}_1)=(2/\pi)\arcsin(\sqrt{\mathcal{T}_1})$ (orange line). Inset: Finite-size scaling for the two-sample Kolmogorov-Smirnov statistic that measures the distance between the empirical cumulative distributions of $\mathcal{T_1}$ and samples of $\mathcal{T}$ drawn from a reference arcsine distribution. Here the black lines are a guide to the eye.
\label{fig:9} }
\end{figure*}

Using the time traces of the stochastic work $W(t)$ we determine the empirical average $\langle W(t)\rangle =N_{\rm cycle}^{-1}\sum_{i=1}^{N_{\rm cycle}} W_i(t)$ with $N_{\rm cycle}=t_{f}/\tau$ given by the total number of cycles of period $\tau$. We use the empirical value of $\langle W(t)\rangle$ to measure  the fraction of time   $\mathcal{T}_1$ that the  work  stays above its average value. This procedure is done for different values of the total observation time $t_f$ under different experimental conditions. Figure~\ref{fig:9} shows that the cumulative distribution of $\mathcal{T}_1$ tends to the arcsine distribution $F(\mathcal{T}_1)=(2/\pi)\arcsin(\sqrt{\mathcal{T}_1})$ for large observation time $t_f$ for several values of $\tau$.

\end{document}